\documentclass[11pt]{article}
\usepackage{moriond,epsfig}

\def\be{\begin{equation}}
\def\ee{\end{equation}}
\def\bea{\begin{eqnarray}}
\def\eea{\end{eqnarray}}

\begin{document}
\hfill DESY 00--046

\vspace*{1cm}
\title{Inflation models, spectral index and observational constraints.}

\author{L. COVI}

\address{DESY Theory Group, Notkestrasse 85\\
22603 Hamburg, Germany}  

\maketitle\abstracts{
We have evaluated the observational constraints on the spectral
index $n$, in the context of a $\Lambda$CDM model.
For $n$ scale-independent, as predicted
by most models of inflation, present data require 
$n\simeq 1.0 \pm 0.1$ at the 2-$\sigma$ level.
We have also studied the two-parameter scale-dependent spectral
index, predicted by  running-mass inflation models. Present data
allow significant variation of $n$ in this case, within the
theoretically preferred region of parameter space.}

\section{Introduction}

It is generally supposed that structure in the Universe originates
from a primordial gaussian curvature perturbation, generated by 
the quantum fluctuations of the inflaton field during slow-roll 
inflation. Then the spectrum of the curvature perturbation
$\delta_H (k)$ is determined by the inflaton potential $V(\phi)$.
In this paper we will consider the scale--dependence of the
primordial spectrum, defined by the spectral index $n$:
\be
n(k)-1\equiv 2 {\partial \ln \delta_H\over \partial \ln k}
=  2M_{Pl}^2 (V''/V)-3M_{Pl}^2 (V'/V)^2 
\, ,
\ee
where\footnote{$M_{Pl}=2.4\times 10^{18}\mbox{GeV}$ is the Planck mass, 
$a$ is the scale factor and $H=\dot a/a$ is the Hubble parameter, and
$k/a$ is the wavenumber.} the potential and its derivatives are 
evaluated at the epoch of horizon exit $k=aH$. 
The value of $\phi$ at this epoch is given by $ \ln(k_{end}/k)=N(\phi)
=M_{Pl}^{-2}\int^\phi_{\phi_{end}} (V/V') d\phi\,$ ,
where $k_{end}$ is the scale leaving the horizon at the end of slow roll
inflation and $N(\phi)$ is the number of e-folds.
In the majority of the inflation models, $n$ is practically 
scale-independent so that $\delta_H^2\propto k^{n-1}$, but
we shall also discuss an interesting class of models giving
significant scale dependence.

\section{The observational constraints on the
$\Lambda$CDM model}

In the interest of simplicity and due to present observations
\cite{lcovi-1}, we adopt the $\Lambda$CDM model, with $\Omega_{tot} =1$ 
and cold non-baryonic dark matter with negligible interaction.
We shall constrain this model, including the spectral index, 
by performing a least-squares fit to the key observational 
quantities. 

The parameters of the $\Lambda$CDM model are the primordial
spectrum $\delta_H(k)$, the Hubble constant $h$ (in units of 
$100\,\mbox{km}\,\mbox{s}^{-1}\,\mbox{Mpc}^{-1}$),
 the total matter density $\Omega_0$,  the 
baryon density $\Omega_b$, and the 
reionization redshift $z_R$ (we consider complete and sudden 
reionization).
$z_R$ can be estimated in terms of the other parameters because 
it can be related to the density 
perturbation and the the fraction of collapsed matter $f$ at 
the epoch of reionization,
so we exclude it from the least-squares fit. In the case of the 
constant $n$ models we fix it at a reasonable value ($z_R = 20$),
while in the case of the running mass models
we compute it assuming that reionization occurs when a 
fixed fraction of the matter $f \simeq 1$ collapses.
The spectrum is conveniently specified by its value at the COBE scale
$k_{COBE}=6.6H_0$, and the spectral index $n(k)$.
Excluding   $z_R$,  the $\Lambda$CDM model is therefore  specified by
five  parameters in the case of a constant spectral index, or by six
parameters in the case of running mass inflation models.

Taking as our starting point a study performed three years ago 
\cite{lcovi-2}, we consider seven  observational quantities: 
the cosmological  quantities $h$, $\Omega_0$, $\Omega_B$, which  
we are also taking as free parameters, and moreover
the shape parameter $\Gamma $, $\sigma_8$, the COBE normalization
and the first peak height in the cmb anisotropy. 
The adopted values and errors are given in the second and third
line of Table 1. For a discussion of the data, see \cite{lcovi-3}. 
In common with earlier investigations, we assume the errors
to be uncorrelated and random errors to dominate over
systematic ones.

\begin{table}[t]
\caption{Fit of the $\Lambda$CDM model to presently available data.
The spectral index $n$ is a parameter of the model, as
are the next four quantities. Every quantity except $n$ is 
a data point, with the value and uncertainty listed in the first 
two rows taken from the references in superscript.
The result of the  least-squares fit is in the
lines three to five for $z_R=20$.  
All uncertainties are at the nominal 1-$\sigma$
level. The total $\chi^2$ is 2.4 for 2 degrees of freedom.}
\begin{center}
\begin{tabular}{|c|c|cccc|ccc|}
\hline
& $n$ & $\Omega_b h^2$ & $\Omega_0$ & $h$ & $10^5\tilde \delta_H$
&$\tilde \Gamma$ & $\tilde \sigma_8$ 
& $\sqrt{\tilde C_{\rm peak}}$ \\[4pt]
data & --- & $0.019\,$\cite{lcovi-4} & $0.35\,$\cite{lcovi-1} 
& $0.65\,$\cite{lcovi-1} &
$1.94\,$\cite{lcovi-5} & $0.23\,$\cite{lcovi-2,lcovi-6} 
& $0.56\,$\cite{lcovi-7} & $80\,\mu\mbox{K}\,$ \cite{lcovi-8}\\[4pt]
error & --- & 0.002 & 0.075 & 0.075 & 0.075 & 0.035 & 0.055 
& $10\,\mu\mbox{K}$
 \\[4pt]
fit & $1.01$ & $0.019$ & $0.36$ & $0.63$
& $1.95$ & 0.19 & 0.58 & $72\,\mu\mbox{K}$ \\[4pt]
error & 0.05 & 0.002 & 0.06 & 0.06 & 0.075 & --- & --- & --- \\[4pt]
$\chi^2$ & --- & $4\times10^{-5}$ & $1\times 10^{-2}$ & 0.1 &
$5\times 10^{-3}$ & $1.3$ & $0.2$ & $0.8$ \\[4pt]
\hline
\end{tabular}
\end{center}
\end{table}

\section{Results}

We perform the least--squares fit with the CERN Minuit package;
the quoted error bars use the parabolic approximation,
while the exact errors computed by Minuit agree with the
approximated ones to better than 10\%.

In order to simplify the numerical procedure, we follow 
\cite{lcovi-9} and parameterize the predicted value of 
$\sqrt{\tilde C_{\rm peak}}$ with the analytical formula 
$ \sqrt{\tilde C_{\rm peak}} = 77.5 \,\mu\mbox{K} 
\left(\frac{\delta_H(k_{COBE})}
{1.94\times 10^{-5}}\right) \left(\frac{220}{10}\right)^{\nu/2} $
where
\be
\nu \equiv 0.88(n_{COBE}-1)-0.37 \ln(h/0.65)-0.16 \ln(\Omega_0/0.35)
+ 5.4 h^2(\Omega_b - 0.019) -0.65 g(\tau)\tau
\,
\end{equation}
and $ \tau= 0.035\frac{\Omega_b}{\Omega_0} h 
\left( \sqrt{\Omega_0(1+z_R)^3 +1-\Omega_0} -1 \right)$.
The formula is fitted to the CMBfast\cite{lcovi-10} results and agrees
within 10\% for a 1-$\sigma$ variation of the 
cosmological parameters, $h, \Omega_0$ and $ \Omega_b$,
and $n=1.0\pm 0.05$.  
With the function $g(\tau)$ set equal to 1, the formula
contains the usual  factor $\exp(-\tau)$. 
By fitting the output of CMBfast, we introduce also $g(\tau)=
1- 0.165 \tau/(0.4+\tau)$, so that our formula is
accurate to a few percent over the interesting 
range of $\tau$.

\smallskip 
{\bf Constant spectral index.}
For the case of a constant spectral index our result is given
in Table 1 for $z_R = 20$. In the case of no reionization
($z_R=0$) we obtain a slightly smaller spectral index,
$n=0.98\pm .05$, and cosmological parameters within the observational
error bar, in agreement with previous analysis \cite{lcovi-11}.
This result is not enough yet to exclude completely proposed 
inflationary models, but
a better determination of the peak height could strengthen the
bound sufficiently to discriminate between them, especially in the
case of new inflation models, which give low values of $n$ \cite{lcovi-3}.

\smallskip
{\bf Running mass models.}
We have also considered the scale-dependent spectral index,
 predicted in inflation models with a running inflaton mass
 \cite{lcovi-12}. In these models, one--loop
corrections to the potential are taken into account by evaluating 
the scale dependent inflaton mass $m^2(Q)$ at $Q\simeq \phi $.
Then the spectral index can be parameterized by just two quantities:
\begin{equation}
{n(k)-1\over 2} = \sigma e^{-c {N(\phi)}} - c
\, ,
\label{runpred}
\end{equation}
where $\sigma$ is an integration constant and $c$ is related to the
inflaton coupling responsible of the mass running. The different signs
of $\sigma $ and $c$ give raise to four different models of inflation.
In general, without fine tuning, we expect
\bea
|c|\mathrel{\rlap{\lower4pt\hbox{\hskip1pt$\sim$}}\raise1pt\hbox{$<$}} 
|\sigma| \mathrel{\rlap{\lower4pt\hbox{\hskip1pt$\sim$}}\raise1pt\hbox{$<$}}
 1 & & |c| \simeq g^2 {\tilde m^2 M_{Pl}\over V_0}
\,
\eea
with $g$ denoting the gauge or Yukawa coupling of the inflaton, 
$\tilde m^2$ the soft 
supersymmetry breaking mass of the particles in the loop and $V_0$ 
the value of the potential energy during inflation.
With gravity-mediated susy breaking, typical values of the 
 masses are $\tilde m^2\sim V_0/M_{Pl}^2$,
which makes $c$ of order of the coupling strength. 
For a gauge coupling, or an unsuppressed Yukawa coupling, we expect
$ |c|\sim 10^{-1}\mbox{ to }10^{-2}$.

Extremizing with 
respect to all other parameters, we have computed the 
$\chi^2$ in the $\sigma$ vs. $c$ plane and obtained contour levels
for $\chi^2$ corresponding to the 70\% and 95\% confidence level in 
two variables. The results are shown in Figure 1.

\begin{figure}
\centering
\psfig{figure=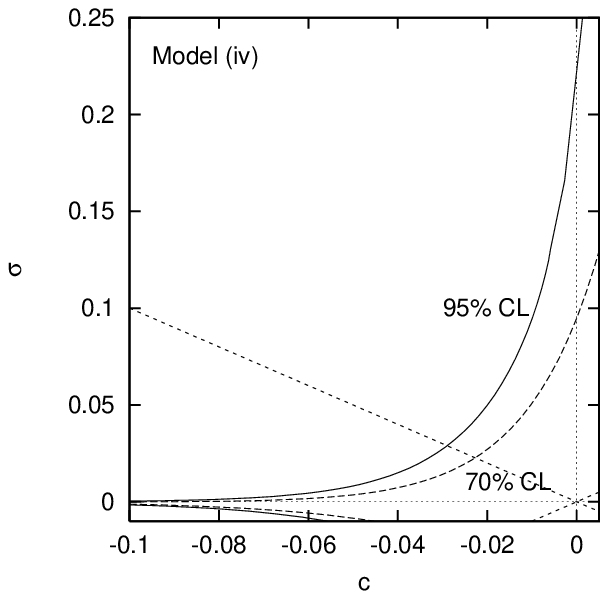,height=6.5cm,width=6.5cm}
\psfig{figure=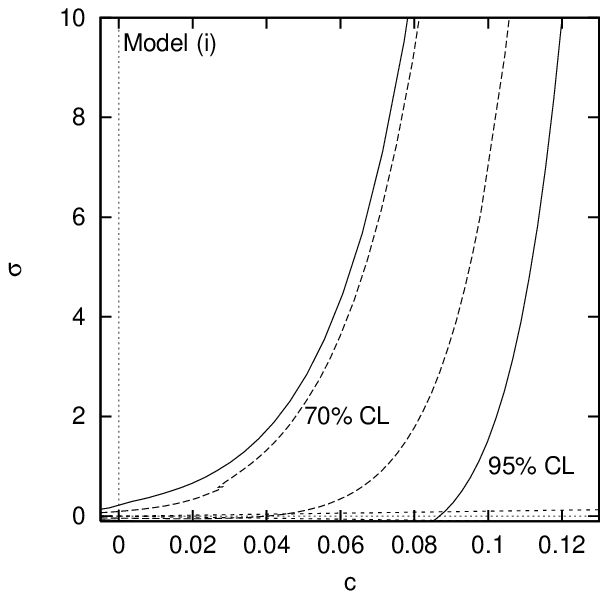,height=6.5cm,width=6.5cm}\\
\psfig{figure=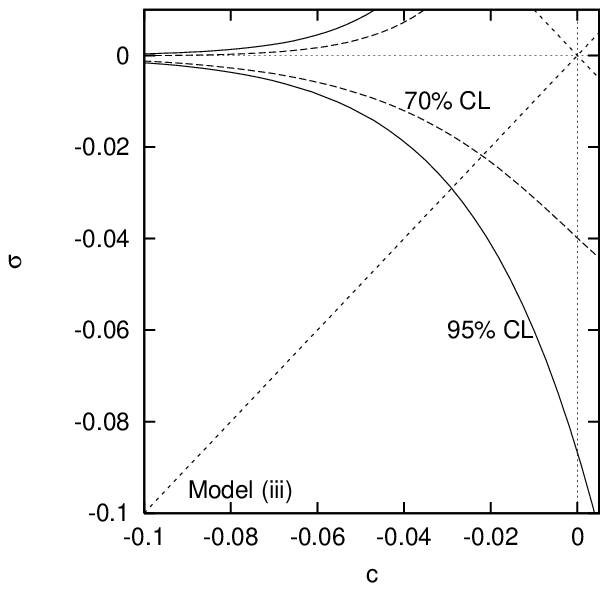,height=6.5cm,width=6.5cm}
\psfig{figure=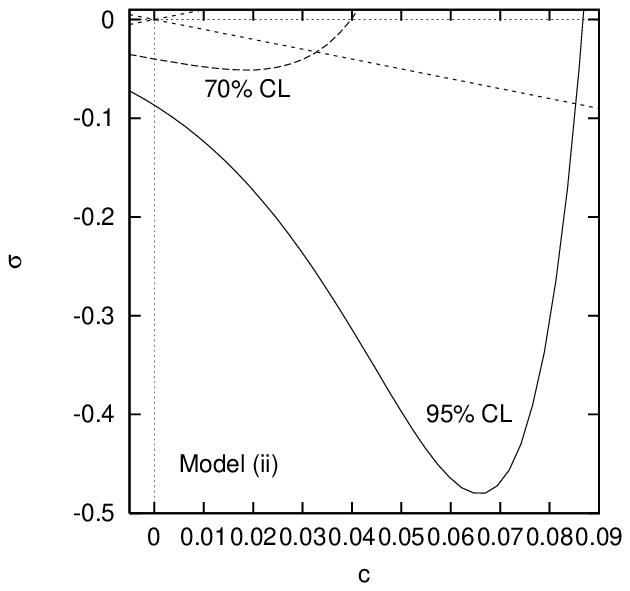,height=6.5cm,width=6.5cm}
\caption[sc-fig1]{Allowed region in the $\sigma$ vs $c$ plane
for $N_{COBE}=50$; the solid line is the 95\% CL contour, 
while the dashed line the 70\% CL. The theoretically favoured region is 
above (below) the dotted line $|\sigma|=|c|$ in the upper (lower) 
half plane. For positive $c$ and $\sigma $ the contours close at 
$\sigma\sim 200, c \simeq 0.15$ and $\sigma\sim 1250, c \simeq 0.19$, 
for the 70\% and 95\% CL respectively.}
\end{figure}

In the case of Models (ii) and (iv), the allowed region corresponds to 
$|c|$ and $|\sigma|$ both small, giving a practically scale-independent
spectral index, with a red and blue spectrum respectively. 

In contrast, the allowed region for Models (i) and (iii) allows 
strong scale-dependence. In Model (i), a large departure from a 
constant spectral index is allowed for large $\sigma $; for the 
theoretically favored value $\sigma\sim 1$ the variation between 
$k_{COBE}$ and $8^{-1} h \mbox{Mpc}^{-1}$ can be as large as 0.05, 
while the maximal change allowed by the data is 0.2. 
For Model (iii), a much larger departure from a constant 
spectral index is allowed, but in  the theoretically favored regime 
$|\sigma|\geq c$ one again finds a variation of at most 0.05.

\section{Conclusion}

We have evaluated the observational constraints on the  spectral index $n$,
using a range of data, and we find, for constant $n$ at 2-$\sigma$ level,
$0.88\leq n\leq 1.11$ for $0 \leq z_R \leq 20$.

We have also investigated the running mass models, parameterized
by $c$ and $\sigma$. 
For $c$ and $\sigma$ with the same sign, we have found that 
indeed $n$ can vary by about $0.05$ between the COBE scale and
$8 h^{-1} \mbox{Mpc}$. Moreover, if $c$ is positive as it would be 
for a gauge coupling, $n-1$ {\em can change sign between the
 COBE and $8h^{-1}\,\mbox{Mpc}$ scales}.
It will be very interesting to see how the present situation changes
with the advent of better data.

\section*{Acknowledgments}

It is a pleasure to thank D. H. Lyth with whom this work has been done.
I would also like to thank the organizers of Moriond 2000 and the 
European Union for the financial support.

\end{document}